\documentclass[conference]{sig-alternate-05-2015}
\usepackage[utf8]{inputenc}
\usepackage{multirow}
\usepackage{graphicx}
\usepackage{color}
\usepackage{xcolor}
\usepackage{listings}
\usepackage{url}
\usepackage{amsmath}
\usepackage{amsfonts}
\usepackage{nomencl}
\usepackage{tikz}
\usepackage{framed}
\usepackage[utf8]{inputenc}
\usepackage[T1]{fontenc}
\usepackage{float}
\usepackage{libertine}
\usepackage{subcaption}
\usepackage{cite}

\begin{document}
\CopyrightYear{2016}
\setcopyright{acmcopyright}
\conferenceinfo{SoICT '16,}{December 08-09, 2016, Ho Chi Minh City, Viet Nam}
\isbn{978-1-4503-4815-7/16/12}\acmPrice{\$15.00}
\doi{http://dx.doi.org/10.1145/3011077.3011121}
\title{Robust Consensus-Based Network Intrusion Detection in Presence of Byzantine Attacks}
\numberofauthors{4}
\author{
\alignauthor
Michel Toulouse\\
       \affaddr{Vietnamese-German University}\\
       \affaddr{Ho Chi Minh City, Vietnam}\\
       \email{michel.toulouse@vgu.edu.vn}	
						\alignauthor Hai Le\\
       \affaddr{Vietnamese-German University}\\
       \affaddr{Ho Chi Minh City, Vietnam}\\
       \email{hai.lh@vgu.edu.vn}
\alignauthor
Cao Vien Phung\\
       \affaddr{Vietnamese-German University}\\
       \affaddr{Ho Chi Minh City, Vietnam}\\
       \email{caovienphung@gmail.com}			\and									
\alignauthor
Denis Hock\\
       \affaddr{Frankfurt University of Applied Sciences}\\
       \affaddr{Frankfurt am Main, Hessen, Germany}\\
       \email{dehock@fb2.fra-uas.de}		
						}


\maketitle

\begin{abstract}
Consensus algorithms provide strategies to solve problems in a distributed system with the added constraint that data can only be shared between adjacent computing nodes. 
We find these algorithms in applications for wireless and sensor networks, spectrum sensing for cognitive radio, even for some IoT services.
However, consensus-based applications are not resilient to compromised nodes sending falsified data to their neighbors, i.e. they can be the target of Byzantine attacks. Several solutions have been proposed in the literature  inspired from reputation based systems, outlier detection or model-based fault detection techniques in process control.  We have reviewed some of these solutions, and propose two mitigation techniques to protect the consensus-based Network Intrusion Detection System in \cite{toulouse2015consensus}. We analyze  several implementation issues such as computational overhead,  fine tuning of the solution parameters, impacts on the convergence of the consensus phase, accuracy of the intrusion detection system. 
\end{abstract}



\section{Introduction}
With the advent of ubiquitous computing on IP-based networks, the
research on appropriate techniques to uncover yet unknown network
misuse patterns and malware gained a great deal of interest.  While a
wide range of different systems has been implemented, many of these
are centralized and show a single point of failure.  The distributed
nature of todays communication networks substantiate the desire to
employ peer-to-peer information exchange in order to reach a global
decision.

In consensus based application schemes, each peer
communicates only with its neighbors and updates its own state
according to an update rule and local weighting.  However, since the
update rules are not resilient against malicious peers sending
incorrect data, the cooperative and fully distributed natures of those
systems expose a high vulnerability to falsification (Byzantine)
attacks \cite{Lamport:1982}.  Different strategies have been proposed in the literature to address this issue based  either on reputation systems \cite{ZengC14}, outlier detection \cite{yan2012vulnerability}, or model based fault detection techniques \cite{Isermann05}. All these strategies have in common that each node acts as an observer of its neighbors and makes a local determination on to whether a neighbor is a compromised  node.

The major contributions of this paper include presenting the impact of
malicious peers on the detection capability of our consensus based Network Intrusion
Detection Systems (NIDS) scheme \cite{toulouse2015consensus}.  We analyze the vulnerabilities of
consensus-based NIDS by proposing data falsification attacks,
which aim to adjust and steal\-thi\-ly manipulate
results.  Moreover, we have implemented two defense strategies to protect the NIDS systems against a single attacker. We compare these strategies under various detection parameters and network topologies
through extensive simulations and analysis using a real IDS and the NSL-KDD
data set \cite{Tavallaee:2009}.

In the remainder, we present an overview of distributed peer-to-peer
NIDS and consensus algorithms as well as related literature.  Next we discuss
the peculiarities of our consensus-based  NIDS.  We point out variations of
falsification attacks and outline two mitigation techniques to adjust the
trustworthiness of participating peers.  Thereafter, we illustrate the
salient features of our prediction model to identify Byzantine peers
and describe a practical experiment we conducted to showcase its
functionality.  Our results demonstrate that, the conducted method can
indeed unveil peers with malicious intend and disruptions in the
information exchange of peer-to-peer NIDS.

\section{Related Work}
Constructing accurate network traffic models with the objective to
discover yet unknown malicious network traffic patterns is, after the
initial publication of Dorothy Denning \cite{denning1987intrusion} and
the implementation of today’s productively used systems such as SnortAD
\cite{szmit2012implementation} and PHAD \cite{mahoney2001phad}, still
a popular field of research.  Methods to subvert these systems, such
as utilizing sophisticated covert channels
\cite{casenove2015exfiltrations} or mechanisms to compromise the
Intrusion Detection System itself \cite{corona2013adversarial} in
order to prevent the detection of malicious activities, have been
proposed to the same extent.

A common mitigation for above mentioned risks is to avoid a single
point of failure by using distributed Intrusion Detection Systems.
Todays alternative to early systems \cite{snapp1991dids,
  bass1999multisensor}, which build upon a master–slave architecture
and require the data to be sent to a central location for analysis,
are peer-to-peer systems \cite{janakiraman2003indra, zhou2005peer}
that recognize attacks in a distributed manner.

Consensus algorithms have a long history in the fields of distributed computing and control theory starting the the work of Degroot \cite{Degroot74}, Chatterjee and Seneta \cite{JPR77} as well as Fisher, Lynch and Paterson \cite{Lynch85}. These initial works have found applications to problems in control theory \cite{1239709}, operations research \cite{4749425} as well as in engineering. One particular domain  of applications concerns  information fusion in distributed systems such as cooperative spectrum sensing in cognitive radio networks \cite{Akyildiz2011}, distributed detection in wireless networks \cite{5306371}, sensor networks \cite{5982199}, services at IoT edge nodes \cite{Li-2014-tii}, and intrusion detection systems \cite{FPVDB08,toulouse2015consensus}. Mitigation techniques against Byzantine attacks on consensus algorithms have been proposed mainly for  cooperative spectrum sensing applications \cite{yan2012vulnerability,6231142,LiuZLLCG12,Yu2009} and  for linear consensus applications \cite{PBB11,5530638,kailkhura2015consensus,5605238}. These mitigation techniques are inspired from outlier detection in machine learning and statistics and model-based fault-detection in control theory.
We have adapted some of those mitigation techniques to protect the consensus phase of a NIDS against Byzantine attacks.

\section{Consensus based NIDS}

A consensus based peer-to-peer NIDS is a set of NIDS modules each placed strategically on a different node of the observed target network (see Figure \ref{fig:top}). An NIDS module consists of traffic sensors that receive copies of all transported packets within the observed network and calculates
an initial local probability for observing benign or malignant network
traffic. Lastly, during the consensus phase, NIDS nodes aggregate their local observations to come-up with a common value which is then used to make a network wide {\it decision} as to whether the network is subjected to malicious activities.

 {\begin{figure}
	\centering
	\includegraphics[width=0.4\textwidth]{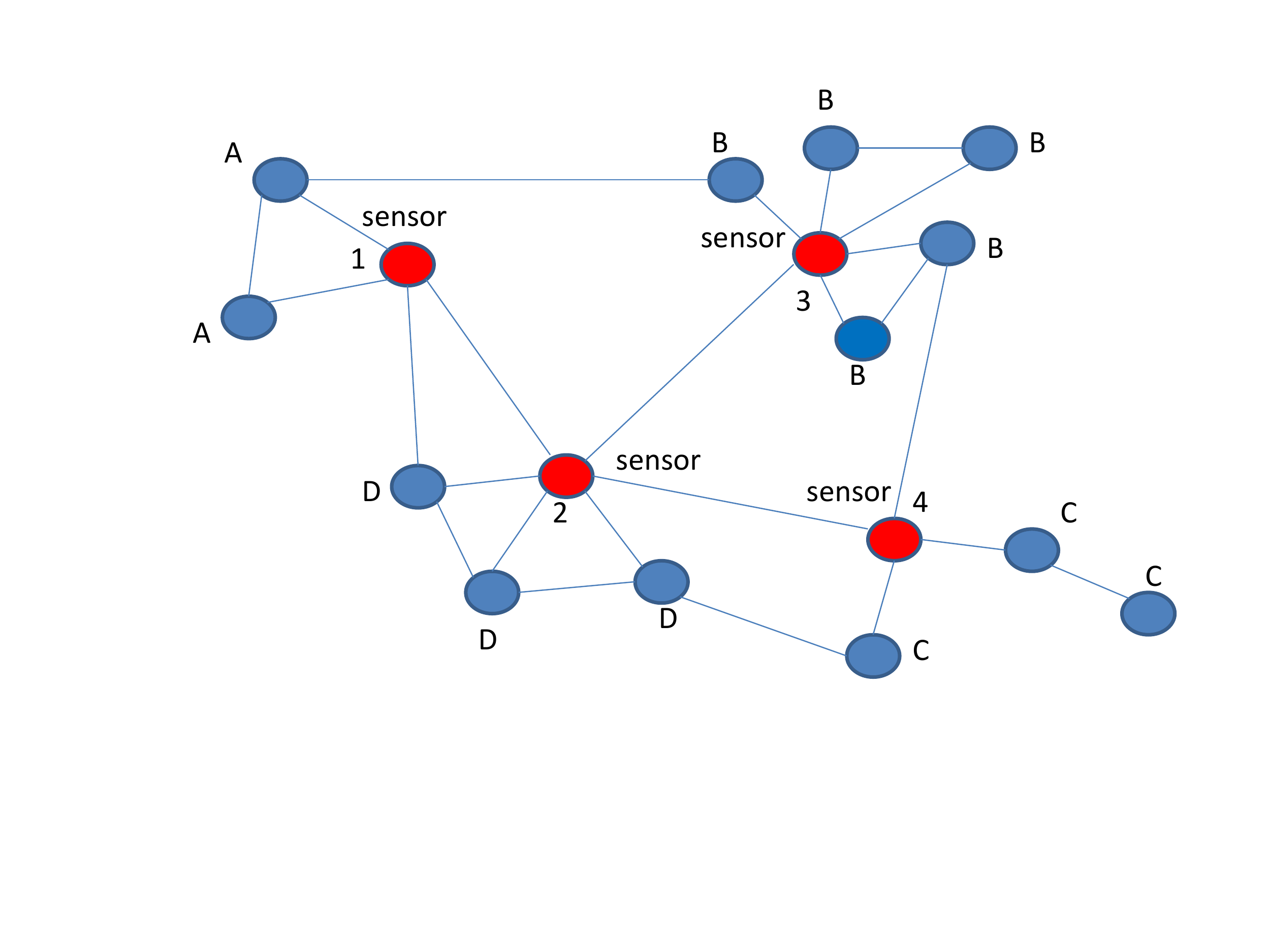} 
	\caption{\label{fig:top} Network Intrusion Detection System.} 
	\end{figure}}
\subsection{NIDS Network Representation}

The NIDS modules observing local network traffic are themselves connected by a network. For the purpose of analysis and comparisons, we study specific topologies of NIDS networks, we refer to such specific network as an {\it NIDS network topology}.  
The topology of a NIDS network is represented by an undirected graph $G = (V, E)$ where $V = \{v_1, v_2, \ldots, v_N\}$ denotes the set of NIDS modules, $E$ denotes the set of edges, we have $(v_i, v_j) \in E$ if and only if there exists a communication link between nodes $v_i$ and $v_j$ in the NIDS network topology. Without lost of generality, we assume that the links connecting pairs of NIDS modules are direct physical links.  It is also assumed that this network is connected, though it does not fully connected pairwise all the NIDS modules.  The {\it adjacency matrix} $A$ of the graph $G$ is such that $a_{ij} = 1$ if and only if $(v_i, v_j) \in E$, $a_{ij} = 0$ otherwise. The {\it neighborhood} of a node $i$ is defined as ${\cal N}_i = \{v_j \in V | (v_i, v_j) \in E\}$, for  $i = 1, 2, \ldots, N$. In Figure \ref{fig:top}, the NIDS neighbors ${\cal N}_2$ of node 2 = $\{1, 3, 4\}$, while ${\cal N}_1 = \{2\}$. The degree of a node $v_i \in G$, denoted as $d_i$, is defined as $d_i = \sum_{i=1}^n a_{ij}$, i.e. it is the number of edges in $E$ that have $v_i$ as endpoint. The {\it degree matrix} $D$ of $G$ is the diagonal 
matrix $diag(d_1, d_2, \ldots, d_N)$. 

\subsection{Network Traffic Analysis and Data Fusion}

\subsubsection{Likelihood Function}
The detection method in the NIDS modules is
"anomaly based" using the well-known naive Bayes classifier to detect Distributed
Denial of Service (DDoS) attacks, such as Land-attack, Syn-flood and
UDP-storm.  The naive Bayes classifier assess the statistical normal
behavior - the 'likelihood' of a set of values to occur - with help
of labeled historic data.  The set of features includes most of the
variables offered by the NLS KDD data set, such as the number of
bytes, service, and number of connections.  The probabilities for an
intrusion is computed for each of these $m$ features. $P(o_j|h)$ expresses the likelihood of the occurrence $o_j$ given the
historic anomalous $h_a$ or normal $h_n$ occurrences.  Thus, if
events receive the same values than benign or malignant network
traffic during training, they result in a high probability for those.
Assuming conditional independence of our $m$ features, the joint
likelihood $P(O_i|h)$ of NIDS module $i$ is the product of all feature likelihoods:

\begin{equation}
P(O_i|h) = \prod_{j=1}^{m} P(o_j|h).
\label{eq:by}
\end{equation}
Each NIDS module locally assigns the joint likelihood, indicating the
abnormality of each event. 

\subsubsection{Consensus Phase}

Let $x_i$ be a variable that represents the state of the NIDS module $i$, and let $x_i(0) = \log(P(O_i|h))$ be the initial state of module $i$, 
where $x_i(0)$ is the likelihood for module $i$ to see a certain set of network
features (see 'Likelihood Function').   The purpose of the consensus phase is for each NIDS module to compute the average sum of the $N$ log-likelihoods: $\frac{1}{N} \sum_{i=1}^N x_i(0)$, while communicating only with direct neighbors. The value $\frac{1}{N} \sum_{i=1}^N x_i(0)$ is used by the NIDS modules to come-up collectively with a same decision  about the state of the network wide traffic. 
The average sum is computed individually by each module $i$ as a weighted sum of  $x_i$ and the $x_j$ for $j \in {\cal N}_i$:

\begin{equation}
x_i(t+1) = W_{ii}x_i(t) + \sum_{j \in {\cal N}_i} W_{ij}x_j(t).
\label{eq2}
\end{equation}
$W_{ij}$ is a weight on edge $(v_i, v_j)$ that uniformly in-cooperates the values of each neighbor and exclude
peers which are not connected. 
The iterates of equation (\ref{eq2}) is the {\it consensus loop} while $x_i(t+1)$ is the {\it consensus value} computed by module $i$ at iteration $t+1$. The {\it consensus phase}  is the computation performed by the $N$ consensus loops. The consensus phase is formally defined  by the following computationally equivalent system: 
\begin{equation}
x(t+1) = Wx(t)
\label{eq2.0}
\end{equation}
where $x(t)$ is a vector of $N$ entries denoting the state of the consensus phase at iteration $t$ and $W$ is the {\it weight matrix} or {\it consensus matrix}.  
The entries of $x(t)$ converge to $\frac{1}{N} \sum_{i=1}^N x_i(0)$ as $t \rightarrow \infty$ provided the  weight matrix satisfies some conditions \cite{toulouse2015consensus}. The {\it max-degree} weight matrix meets these requirements, it is used in this paper:

\begin{equation}
W_{ij} = 
\begin{cases}
    \frac{1}{d+1}      & \quad \text{if } j \in \mathcal{N}_i   \\
		1- \frac{d_i}{d+1} & \quad \text{if } i = j\\
    0                             & \quad \text{if } j \notin \mathcal{N}_i\\
  \end{cases}
\label{eq:cons}
\end{equation}
where $d = max(d_i)$. The stopping condition of a consensus loop (also known as 'convergence parameter') is typically $|x_i(t) - x_i(t-1)| < \epsilon$. A consensus phase is completed once each consensus loop has met the stopping condition. The {\it number of iterations} of a consensus phase is given by the consensus loop that has the largest number of iterations. The {\it convergence speed} of a consensus phase is the number of iterations needed by a consensus phase to complete. The value of $\epsilon$ is set such to minimize the number of iterations during the consensus phase while insuring accuracy of the decision about the state of the network traffic.

\section{Data Falsification Attacks and Mitigation Techniques}

NIDS as well as their sensors are part of the
affected target infrastructures, and thus may become targets of
attacks which
aim to circumvent or degrade detection capabilities.  While
centralized NIDS may become an information bottleneck and undermine
system performance, peer-to-peer NIDS harbor vulnerabilities in their
information exchange.

\subsection{Data Falsification Attacks}

Data falsification attacks ultimately try to mask intrusive traffic to the  consensus-based 
NIDS by degrading the NIDS system accuracy defined as
$\frac{TP + TN}{TP + TN + FP +FN}$,
where  $TP$ ({\it  True Positive}) is the number of attacks detected when it is actually an attack;
$TN$  ({\it True Negative}) is the number of normals detected when it is actually normal;
$FP$ ({\it False Positive}) is the number of attacks detected when it is actually normal; $FN$ ({\it False Negative}) is the number of normals detected when it is actually an attack. Accuracy is reduced  by increasing the number of false positives or false negatives. To mask currently occurring malicious traffic, data falsification attacks could reduce the true probability of attack, here increasing the number of false negatives. Attackers may also plan for long term by  increasing the number of false positives thus reducing the reliability of the system in the eyes of the system administrators.

\begin{figure}[]
  \begin{center}
    \includegraphics[width=1\linewidth]{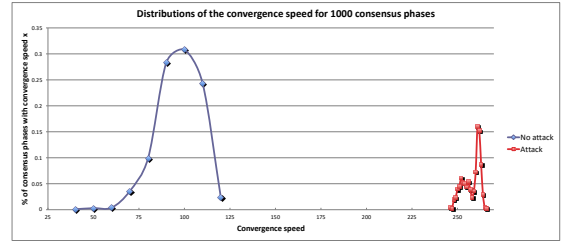}
		\vspace{-15mm}
    \caption{\label{fig2} Convergence speeds with and without loop disruption.}
  \end{center}
\end{figure}

Data falsification attacks on consensus-based detection algorithms can take the following forms \cite{kailkhura2015consensus}: 1- sensor values are falsified, thus the consensus loop is initialized with values originating from falsified network traffic readings; 2- the consensus loop iterations are disrupted. In consensus loop disruption, the simplest case is where attackers ignore the consensus value computed at each iteration and keeps transmitting the same constant $c$. Figure \ref{fig2} illustrates the impact of this attack strategy on the convergence speed of the consensus phase. 

Figure \ref{fig2} plots the distribution of the convergence speed of 1000 consensus phases each having only honest NIDS modules versus a situation where each consensus phase has one compromised module that send the same constant value $c$ to its neighbors. 
Figure \ref{fig2} shows that convergence speed is much slower in the compromised system. In this case, modules all converge to $c$ rather than $\frac{1}{N} \sum_{i=1}^N x_i(0)$.

In a second form of consensus loop disruption, the compromised module send to its neighbors a falsified consensus value. This form of loop disruption is modeled in equation (\ref{eq2.2}), which modifies the consensus loop in (\ref{eq2}) to incorporate this form of attacks:

\begin{equation}
x_i(t+1) = W_{ii}x_i(t) + \sum_{j \in {\cal N}_i} W_{ij}x_j(t) + u_i(t).
\label{eq2.2}
\end{equation}
The consensus value $x_i(t+1)$ is falsfied by adding $u_i(t)$ at each iteration $t$ of the consensus loop. The two mitigation techniques proposed in this paper address this form of loop disruption attack.   
Other attack models, including multiple colluding attackers, are described in \cite{yan2012vulnerability,s16020252}.

\subsection{Mitigation Techniques}

We propose two mitigation techniques that handle consensus loop disruptions by a single Byzantine attacker. The first one is an outlier detection procedure associates with each module to evaluate at each consensus loop iteration the potential that a neighbor of the module is compromised. The second one is an adaptation to cyber-attacks of a model-based fault-detection technique in process engineering and control theory.

\subsubsection{Outlier Detection}

Outlier detection techniques are commonly applied to detect data falsification attacks in wireless sensor networks \cite{Illiano2015}. These techniques use distance thresholds between the value $x_j(t)$ sent by a neighbor $j$ to node $i$ and some reference value $r_i$. For example, if $r_i(t) = x_i(t)$, neighbor $j$ is flagged as intruder if  $|x_j(t) -x_i(t)| > \lambda$ for some threshold value $\lambda$. A unique predefined threshold for all nodes may easily be discovered by an intruder. Furthermore, absolute differences $|x_j(t) -x_i(t)|$ converge to zero as $t \rightarrow \infty$, rendering the outlier detection potentially insensitive when the absolute differences get smaller than $\lambda$.

{\it Adaptive thresholds} have been proposed to address the above issues. It consists for each node to compute a localized threshold, adapting the threshold at each consensus iteration to the reduction of absolute differences $|x_j(t) -x_i(t)|$.
In \cite{LiuZLLCG12}, the threshold \begin{equation}\lambda_i(t+1) = \frac{\sum_{j \in {\cal N}_i}|x_j(t+1) - x_i(t+1)|}{\sum_{j \in {\cal N}_i}|x_j(t) - x_i(t)|} \lambda_i(t) \label{r4}\end{equation} (for properly initialized $\lambda_i(0)$) is computed by each node $i$ and at each iteration of the consensus phase.  $\lambda_i(t)$ partitions neighbors of node $i$ into two sets, those neighbors $j$ that have a deviation $|x_j(t) - x_i(t)| \ge \lambda_i(t)$ are considered suspicious, they constitute the neighborhood ${\cal N}_i^F$ of states that have less weight in the computation of the consensus value $x_i(t+1)$:

$$x_i(t+1) = x_i(t) + \epsilon \sum_{j \in {\cal N}_i^T} x_j(t) + \frac{\epsilon}{a} \sum_{j \in {\cal N}_i^F} x_j(t)$$
for some constant $a$. In adaptive local threshold techniques, thresholds are computed using the diffusion dynamics of consensus algorithms, unless attackers can get multi-hops information access, they cannot foresee the value of their neighbor thresholds, therefore they cannot adapt their data falsification function to keep the values under the radar of the detection procedure. Finally, as the network converge towards consensus, the value $\lambda$ converges toward zero, leading to the attackers to be eventually filter out.

The outlier detection  method we use in this paper computes the threshold $\lambda$ as in equation (\ref{r4}). 
Those neighbors $j$ that have a deviation $|x_j(t) - x_i(t)| \ge \lambda_i(t)$ are flagged as suspicious.  We use a majority rule similar to \cite{yan2012vulnerability} to convert the status of a neighbor NIDS module $j$ from suspicious to attacker. Let $B$ be the number of common neighbors between module $i$ and module $j$. If more then $\lceil\frac{B}{2}\rceil$ neighbors of module $i$ report $j$ as suspicious then module $j$ is considered as an attacker, it is disconnected from the intrusion detection system. Note that we assume a single attacker, if the majority rule identifies more than one neighbor as attacker, the one with the largest deviation is disconnected.

\subsubsection{Model-based Fault Detection}

Fault detection is a field of control engineering concerned with identifying and locating faults in a system. 
Methods in this field essentially compare measurements of the actual behavior of a system with its anticipated behavior. In {\it model-based} fault detection,  the anticipated behavior is described using mathematical models \cite{Isermann05}, the measured system variables are compared with their model estimates.  Comparisons between the system and the model show deviations when there is a fault in the real system. Such difference between the system and its model is called {\it residual} or {\it residual vector}.
There exist several implementations of the model-based approach, the {\it observer-based technique} \cite{Patton96} 1- seeks to discriminate between deviations caused by faults in the real process from those caused by the estimations (which is why it is also called a "filter"); 2- provides a residual vector that indicates the faulty system component (so called directional residual). Observer-based fault detection approaches have been applied to detect cyber-attacks \cite{5530638,PBB11,5605238}.

To detect Byzantine attacks on the intrusion detection system, a loop parallel to the consensus loop, the {\it observer}, is added to each NIDS module. 
The observer loop computes a state vector $x^o(t)$ estimating $x(t)$.
The definition of the observer requires input from the consensus loop. First, we model the consensus loop disruption attack of equation (\ref{eq2.2}) in matrix form:

\begin{equation}
x(t+1) = Wx(t) + {\bf I}_Nu(t)
\label{eq4}
\end{equation}
where ${\bf I}_N$ is the $N$-dimension identity matrix, and where $u_i(t) = 0$ whenever module $i$ behave normally. The consensus loop for a given module $i$ is now defined as follow:
\begin{equation}
\begin{aligned}
&x(t+1) = Wx(t) + {\bf I}_Nu(t) \\
&y_i(t) = C_ix(t) 
\end{aligned}
\label{eq6}
\end{equation}
where $C_i$ is a $(deg_i + 1) \times N$ matrix in which entry $C_i[k,l] = 1$ if $l \in {\cal N}_i$, otherwise $C_i[k,l] = 0$. The vector $y_i(t)$ has $(deg_i + 1)$ entries, each entry $j$ of $y_i(t)$ stores the state $x_j(t)$ at time $t$ of modules $j \in {\cal N}_i$. $y_i(t)$  is part of the definition of the observer. Note, each NIDS module $i$ knows the consensus matrix $W$ and the matrix $C_i$. However, the set of non-zero $u_i$ is unknown to the non-malicious modules.

To detect a malicious neighbor of module $i$, an observer of the consensus sub-system (\ref{eq6}) is defined as follow \cite{fp-ab-fb:06v}:

\begin{equation}
\begin{aligned}
&z(t+1) = (W + GC_i)z(t) - Gy_i(t) \\
&x^o(t) = Lz(t) + Ky_i(l)
\end{aligned}
\label{eq7}
\end{equation}
where $z(t)$ is the state of the observer and $x^o(t)$ is the estimation by the observer of module $i$  of the consensus state $x(t)$. 
The matrices to compute $z(t+1)$ and $x^o(t)$ are defined as follow:  $G = -W_{{\cal N}_i}$, $K = C^T_i$, $L = {\bf I}_N - KC_i$, where $W_{{\cal N}_i}$ are the columns of $W$ with indexes in ${\cal N}_i$. Analysis of this system in its full generality  is obviously beyond the scope of this paper, we refer to \cite{Patton96} for an historical development and analysis of observer-based fault detection systems. 
The analysis of  (\ref{eq7}) can be  simplified as the consensus system (\ref{eq6}) satisfies some conditions \cite{PBB11}.  It can be show that as $t \rightarrow \infty$, $x^o(t) \rightarrow x(t)$, consequently the {\it estimation error} $e(t) = x^o(t) - x(t)$  converges to {\bf 0}. We have
\begin{equation}
x^o_j(t) = \left\{ \begin{array}{ll}
       x_j(t) & \mbox{ if } j = i \mbox{ or } j \in {\cal N}_i\\
 z_j(t) & \mbox{otherwise}
   \end{array}
\right.
\label{eq16}
\end{equation}
The state of the observer $z(t+1)$ can be expressed in terms of the consensus matrix \cite{fp-ab-fb:06v}:
 \begin{equation}
z(t+1) = Wx^o(t).
\label{eq14}
\end{equation}
The {\it iteration error} $\varepsilon(t)$:
 $$\varepsilon(t) = |x^o(t+1) - Wx^o(t)|$$
can then be used as residual vector. From (\ref{eq16}) and (\ref{eq14}), $\varepsilon_j(t) = 0$ for $j \not = i$ and $j \not \in {\cal N}_i$. If $\varepsilon_j(t) \not = 0$, either  $x^o_j(t) \not = x_j(t)$ (estimation error is greater than 0), or $u_j \not = 0$. Since the estimation error dissipates as $t \rightarrow \infty$,
we have  $(x^o(t+1) - Wx^o(t)) \rightarrow {\bf I}_Nu(t)$ as $t \rightarrow \infty$. If $u_j \not = 0$ for some $j \in \{1, \ldots, N\}$ then $(x^o_j(t+1) - Wx^o_j(t)) \rightarrow u_j(t)$, the corresponding module $j$ is detected as an Byzantine  attacker.

The observer defined in  (\ref{eq7}) is run in parallel with the iterations of the consensus loop of each module $i$. Together with the consensus loop in (\ref{eq6}), it  provides an algorithm to detect Byzantine attacks at the level of each NIDS. Each module $i$ build a consensus system and an observer as described in equations (\ref{eq6}) and (\ref{eq7}). At each consensus iteration, each module $i$ computes $\varepsilon(t) = |x^o(t+1) - Wx^o(t)|$. 
If $\varepsilon_j(t) \not = 0$ then module $j \in {\cal N}_i$ is compromised, and should be isolated from the other modules of the intrusion detection system. 

\section{Empirical Analysis}

We have implemented the above two mitigation techniques as part of the system described in \cite{toulouse2015consensus}. Tests are performed with four different NIDS network topology simulations: rings, 2-dimensional torus with 9 and 25 NIDS modules, Petersen graph (10 nodes 15 edges) and several random graphs having the same number of vertices and edges as in the Petersen graph (the configuration of an NIDS network, such as topology and number of nodes, impacts some aspects of the consensus phase, we hypothesized  it will be the case as well for the detection of Byzantine attacks). Each test consists of 1000 iterations of the above NIDS networks. In {\it one test iteration}, each NIDS module reads the local network traffic from one entry of the KDD data set, analyzes the local traffic, then executes its consensus loop.   The analysis of the local network traffic  returns two values: $p_a$ the probability that the observed traffic is intrusive; $p_n$  the probability the observed traffic is normal. During the consensus phase, each module compute $\frac{\sum_{i=1}^N \log(p_{a_i})}{N}$ and $\frac{\sum_{i=1}^N \log(p_{n_i})}{N}$. In tests with the outlier method, each module execute the following consensus loop: 
\begin{equation}
x^a_i(t+1) = W_{ii}x^a_i(t) + \sum_{j \in {\cal N}_i} W_{ij}x^a_j(t) + u_i(t)
\label{eq2.23}
\end{equation}
\begin{equation}
x^n_i(t+1) = W_{ii}x^n_i(t) + \sum_{j \in {\cal N}_i} W_{ij}x^n_j(t) + u_i(t).\\
\label{eq2.22}
\end{equation}
In tests with the fault detection method, each module execute the following consensus loop: 
\begin{equation}
\begin{aligned}
&x^a(t + 1) = Wx^a(t) + {\bf I} u^a\\
&y^a_i(t) = C_ix^a(t)
\end{aligned}
\label{eq9}\end{equation}
\begin{equation}
\begin{aligned}
&x^n(t + 1) = Wx^n(t) + {\bf I} u^n\\
&y^n_i(t) = C_ix^n(t).
\end{aligned}
\label{eq10}\end{equation}
The $n$-dimensional vectors $x^a(0)$ and $x^n(0)$ are initialized respectively as $x^a_i(0) = \log(p_{a_i})$ and $x_i^n(0) = \log(p_{n_i})$, for $i = 1 ..n$. The observer loop of the fault detection method is identical to the system described in (\ref{eq7}). In all cases, the decision to raise an alert is based on the ratio of the system wide consensus values $\frac{p_a}{p_n}$ and some predefined alert value. We have filtered attacks in NSL-KDD data set to retain only denial of service attacks.

Attacks inject positive values in the consensus loop component (\ref{eq2.23}) or (\ref{eq9}), thus increasing the number of false positives. At each iteration of a test, a to be compromised  NIDS module $j$ is selected randomly,  $u^a_j$ is then assigned with a positive value.
The magnitude of $u^a_j$ has to be large enough to falsify the decision at the end of the consensus phase (i.e. raise an alert when traffic is normal), if not detected. For example, $u^a_j = 0.0005$ is to small, it does not have an impact on the decision. However, a value such as $u^a_j = 0.5$ can cause the system to converge on a value different enough from $\frac{1}{N} \sum_{i=1}^N x_i(0)$ such to have an impact on the system decision. Having $u^a_j = 0.5$ is also suitable to obtain meaningful test results. The values $p_{a_i}$ and $p_{n_i}$ returned by the Bayesian analysis of a module $i$ are the product of likelihoods $\prod_{j=1}^{m} P(o_j|h)$, as the number of features is large, the product of likelihoods are very small. During the consensus phase, neighbor NIDS modules exchange log-likelihoods, which are in ranges between -20 and -55.  So $u^a_j = 0.5$ is a relativity small external input to the consensus phase. It is large enough so that our two mitigation techniques always detect an attack, but failing to detect it soon enough can lead the consensus phase to converge to values quite different from $\frac{1}{N} \sum_{i=1}^N x_i(0)$.

In the following, we first evaluate the computational cost of running each of the two mitigation techniques. Subsidiary, we also report the number of consensus iterations needed to detect an Byzantine attack. Next we analyze the efficiency of the mitigation techniques to prevent the occurrence of false positives at the conclusion of a consensus phase.

Table \ref{table1} reports the computational cost of running each mitigation technique. The tests are executed while no attack take place. The column "Cost" reports the time in milliseconds for running the NIDS network simulation during 1000 iterations. 
In Table \ref{table1}, rows "no detection" give the cost of running a NIDS network topology without the execution of any detection code. Rows "outlier" and "fault" give the cost of running NIDS modules while also executing respectively the code for the outlier method and the fault method.  
The higher costs of the mitigation techniques compared to "no detection" for the same network size and topology reflects the cost for protecting the NIDS with each of the mitigation techniques. Table \ref{table1} shows that the computational overhead for outlier is clearly less than for the fault detection method.   
\begin{table}[htb]
\caption{Computational cost of the mitigation techniques.}
\label{table1}
\centering
 \begin{tabular}{|c |c| c |c |}
\hline
Topology & Size & Detection & Cost\\
\hline
\multirow{6}{*}{Ring}& \multirow{3}{*}{9} &no detection &0.050 \\\cline{3-4}
&&outlier &0.276\\\cline{3-4}
&&fault &0.921\\\cline{2-4}
& \multirow{3}{*}{25}&no detection & 0.101\\\cline{3-4}
&&outlier &1.131\\\cline{3-4}
&&fault &3.286\\\hline
\multirow{6}{*}{Torus}& \multirow{3}{*}{9} &no detection & 0.027\\\cline{3-4}
&&outlier &0.121\\\cline{3-4}
&&fault &1.327\\\cline{2-4}
& \multirow{3}{*}{25}&no detection &0.043 \\\cline{3-4}
&&outlier &0.567\\\cline{3-4}
&&fault &6.055\\\hline
\multirow{3}{*}{Petersen}& \multirow{3}{*}{10} &no detection &0.005 \\\cline{3-4}
&&outlier &0.135\\\cline{3-4}
&&fault &0.597\\\hline
\multirow{3}{*}{Random}& \multirow{3}{*}{10} &no detection &0.013 \\\cline{3-4}
&&outlier &0.290\\\cline{3-4}
&&fault &1.268\\\hline
 \end{tabular}  
\end{table}

\begin{figure}[!thb]

  \begin{center}
    \includegraphics[width=1\linewidth]{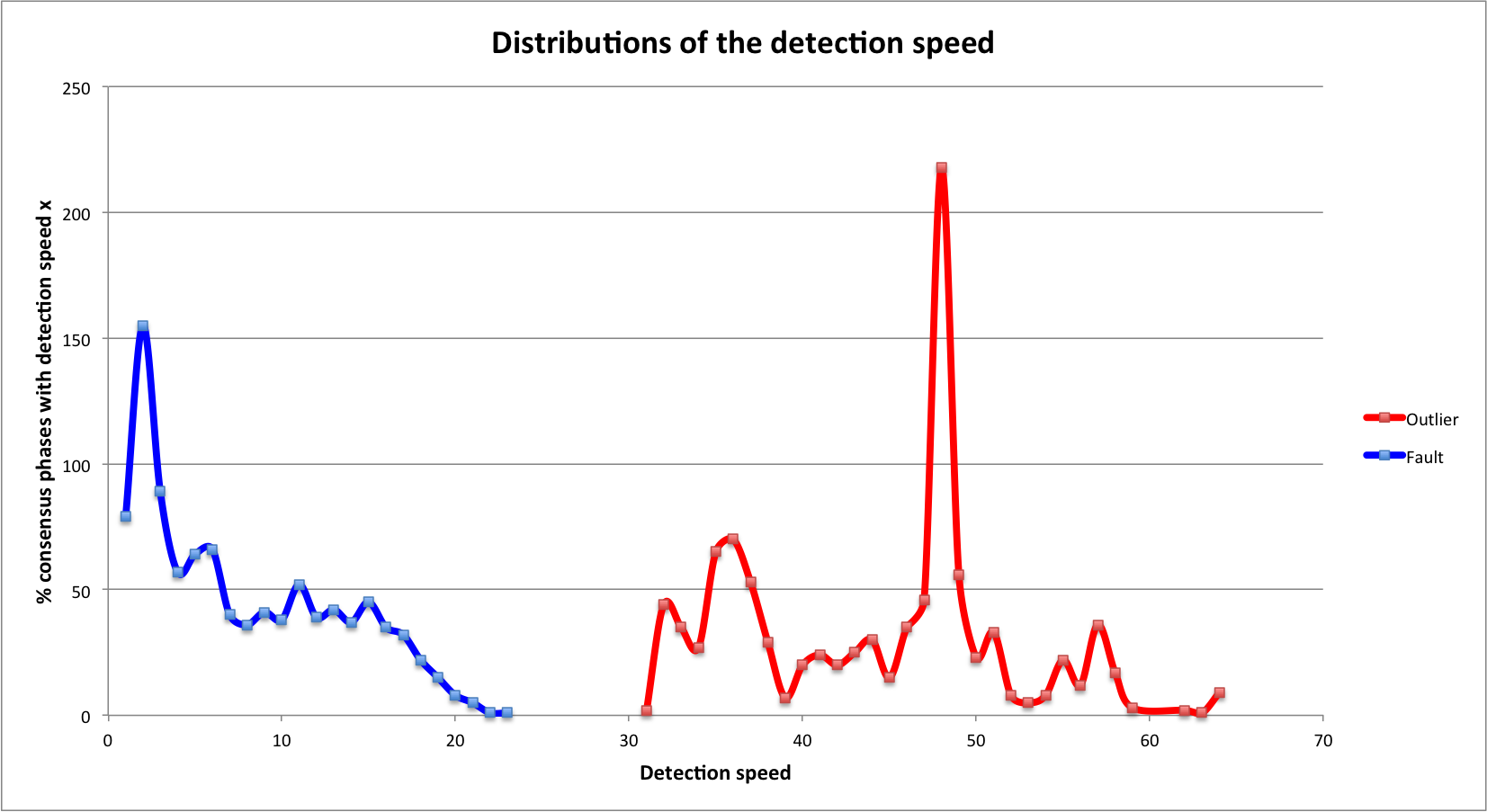}
		
    \caption{\label{fig3} Detection speed of ring topology 9 nodes.}
  \end{center}
\end{figure}

\begin{figure}[!thb]

  \begin{center}
    \includegraphics[width=1\linewidth]{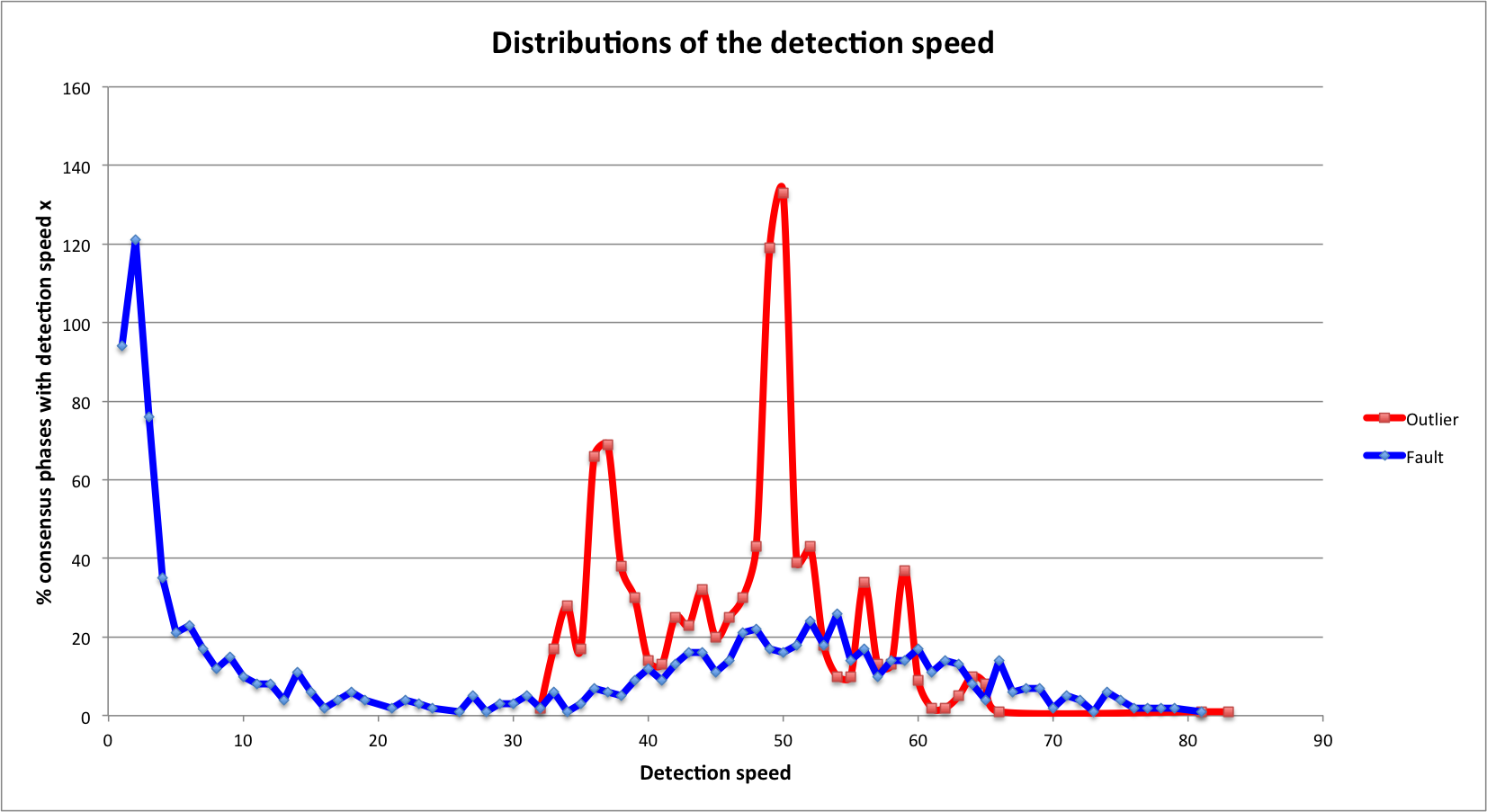}
		
    \caption{\label{fig4} Detection speed of ring topology 25 nodes.}
  \end{center}
\end{figure}

\begin{figure}[!thb]

  \begin{center}
    \includegraphics[width=1\linewidth]{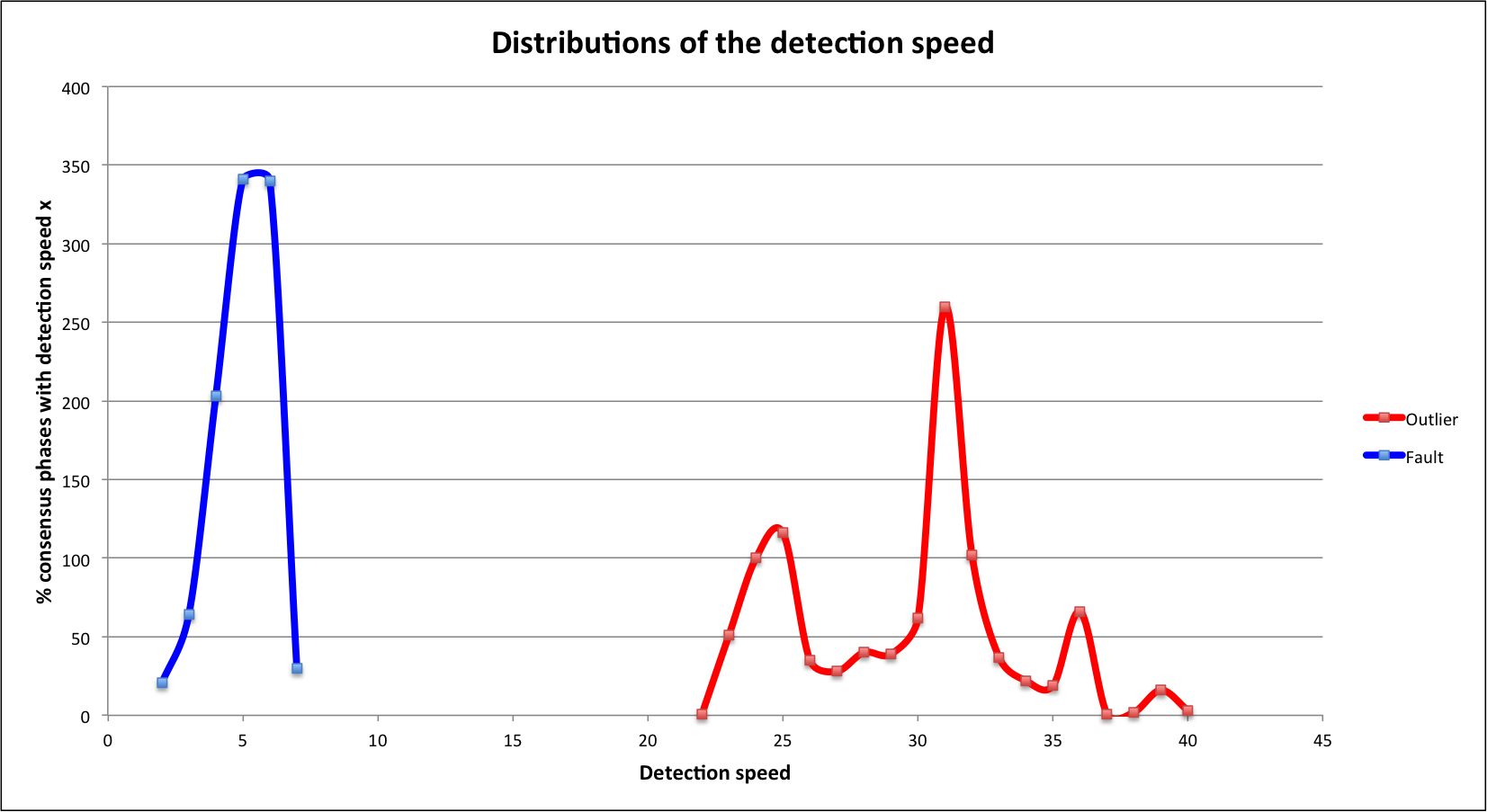}
		
    \caption{\label{fig5} Detection speed of torus topology 9 nodes.}
  \end{center}
\end{figure}

\begin{figure}[!thb]

  \begin{center}
    \includegraphics[width=1\linewidth]{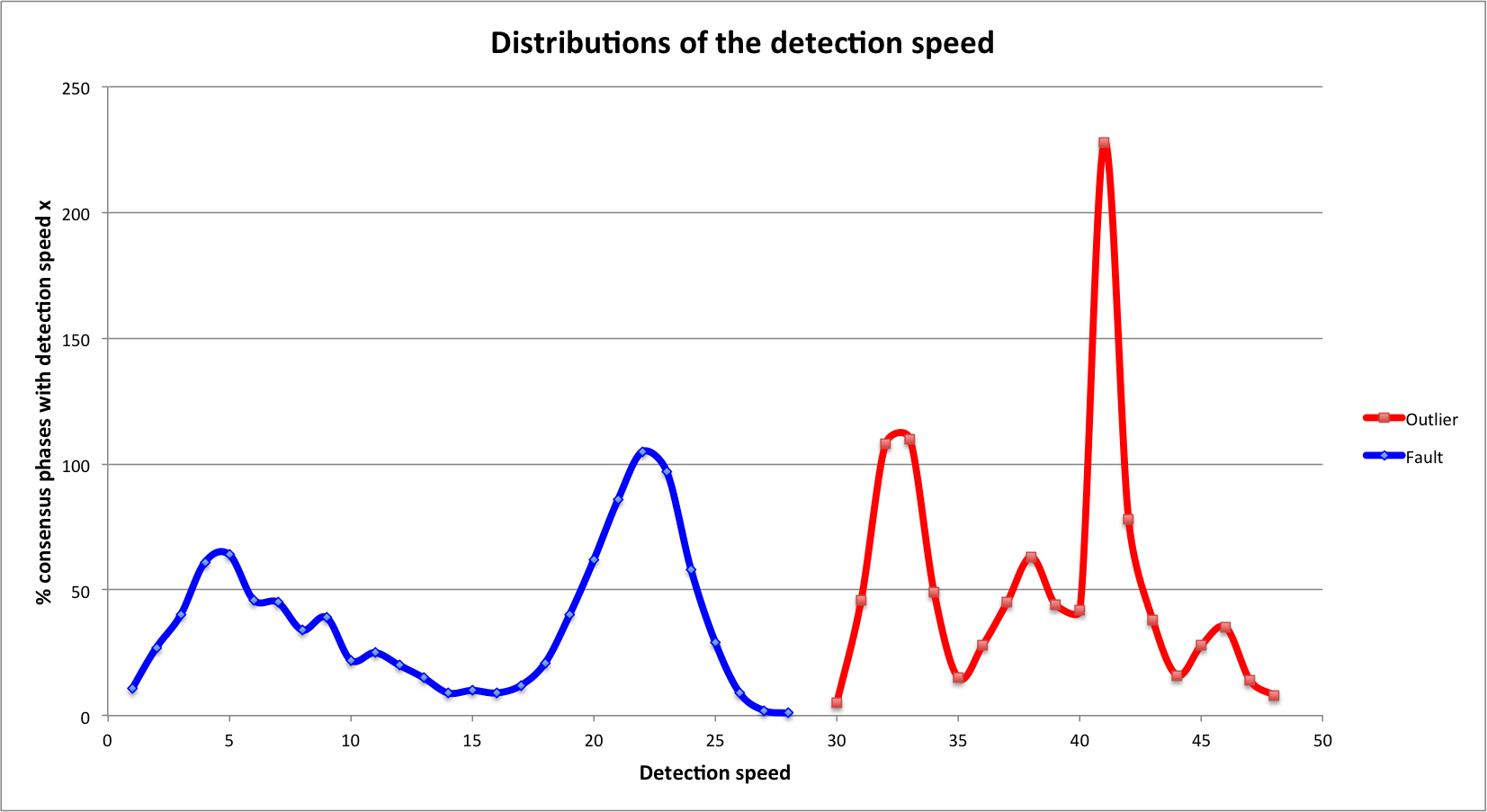}
		
    \caption{\label{fig6} Detection speed of torus topology 25 nodes.}
  \end{center}
\end{figure}

\begin{figure}[!thb]

  \begin{center}
    \includegraphics[width=1\linewidth]{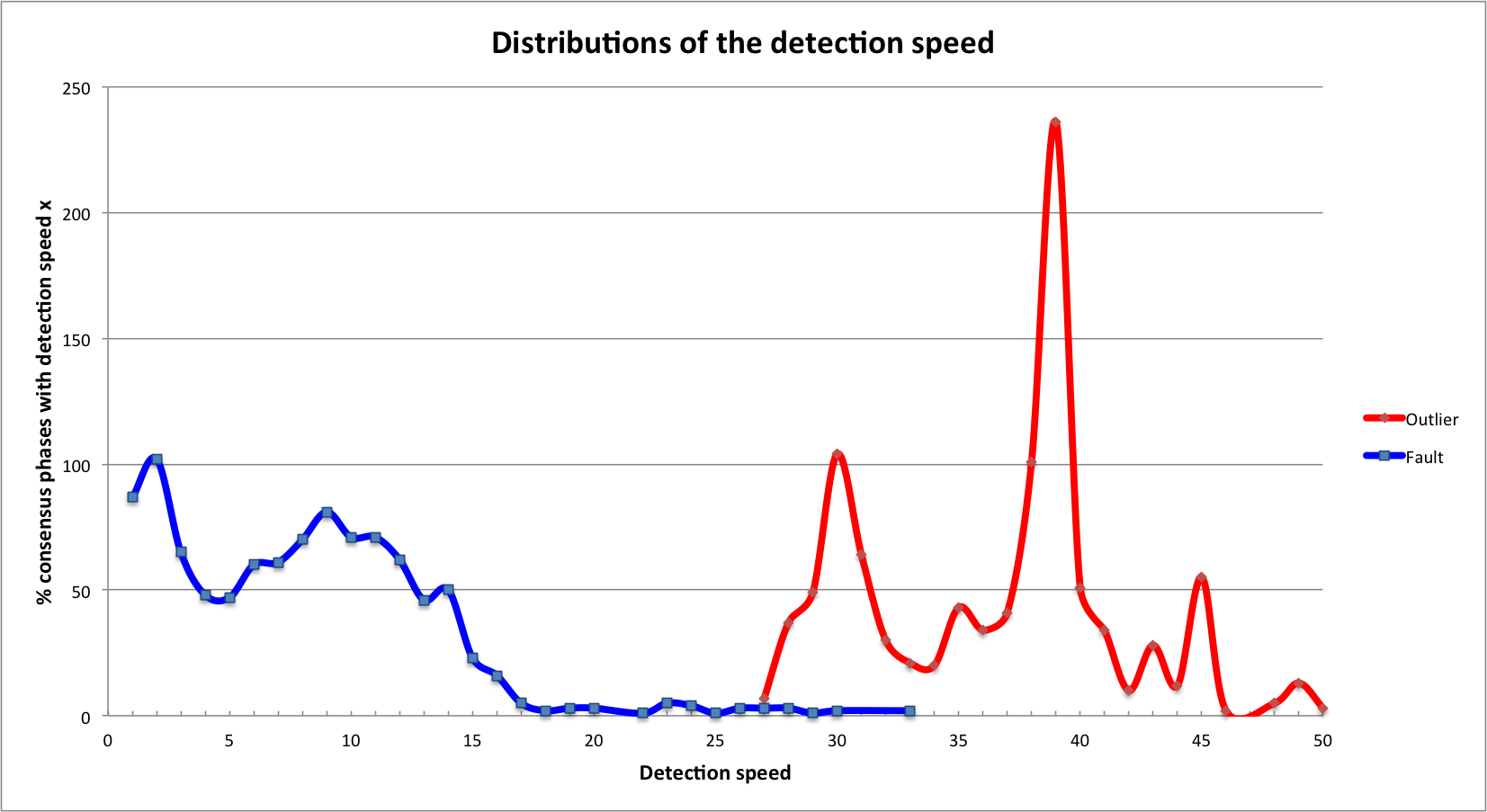}
		
    \caption{\label{fig7} Detection speed of Petersen graph.}
  \end{center}
\end{figure}

\begin{figure}[!thb]

  \begin{center}
    \includegraphics[width=1\linewidth]{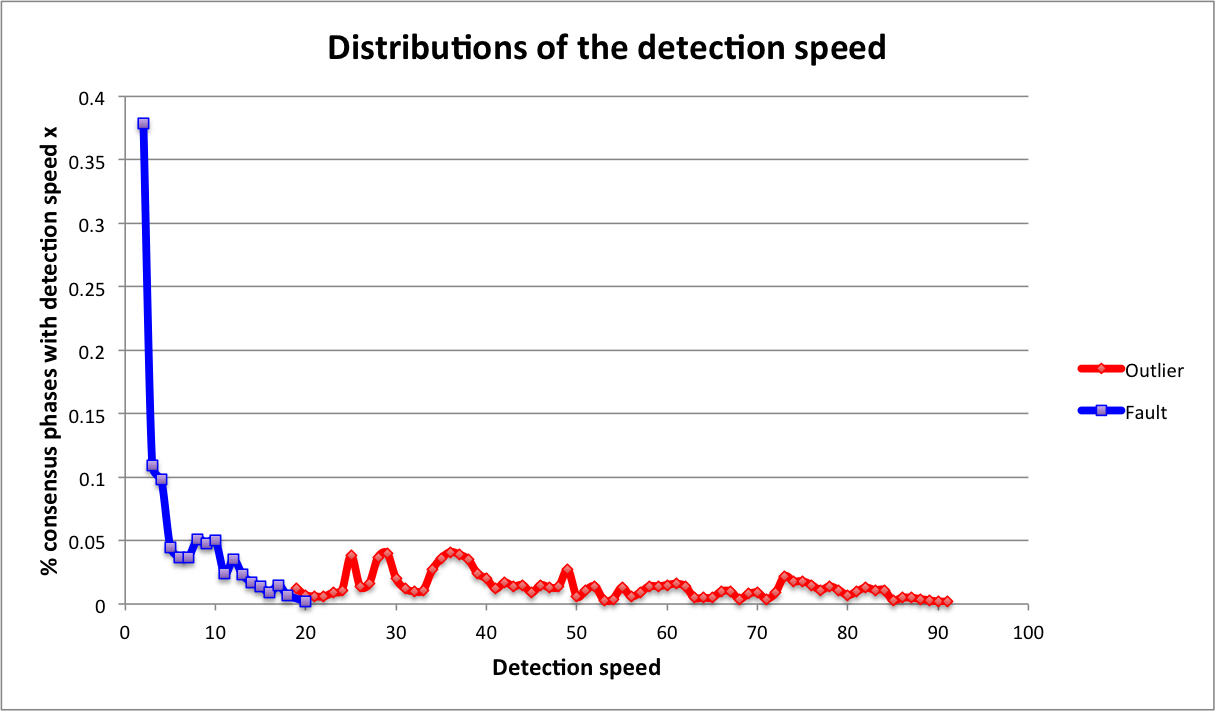}
		
    \caption{\label{fig8} Detection speed of Random graphs.}
  \end{center}
\end{figure}

Figures \ref{fig3} to \ref{fig8} analyze the {\it detection speed} of each mitigation technique.  The detection speed is the number of consensus iterations executed before an Byzantine attack is discovered and eliminated. The analysis is performed for the different network topologies. The values on the $x$ axis are the number of consensus iterations needed to detect the Byzantine attacker. The $y$ axis displays the percentage of the 1000 tests that needed a given number of consensus iterations to detect the attacker.
These figures clearly show that the fault detection approach needs fewer iterations to detect and terminate Byzantine attacks. Comparing with the computational cost in Table \ref{table1}, the outlier method has a more favorable computational overhead but requires more iterations to detect intruders. Note that the detection speed of the fault detection method depends on the capacity to discriminate iteration error $|x^o(t+1) - Wx^0(t)|$ caused by an estimation error $x^o(t) - x(t)$ and caused by external inputs. A minor improvement here will consist of using outlier methods to identify the intruder. Increasing the magnitude of $u^a_j$ beyond 0.5 increases the magnitude of the iteration error vector entry $|x^o(t+1) - Wx^0(t)|$ corresponding to the attacker compared to entries impacted only by estimation errors, which can be detected by an outlier method. Conversely, improving convergence speed of the estimation error $x^o(t) - x(t)$ towards zero can help to flag the attacker as outlier. This will avoid to wait for the convergence of $(x^o(t+1) - Wx^o(t)) \rightarrow {\bf I}_Nu(t)$, where ${\bf I}_Nu(t)$ is the attack vector.



Attacks have an impact on the accuracy of the decision made by the NIDS about the state of the network traffic. Table \ref{table2} measures the effectiveness of the mitigation techniques to maintain the accuracy of the NIDS. The "no attack" rows report the accuracy of the NIDS system when no attack occur. The "no detection" rows report the accuracy of the NIDS when attacks take place  and the NIDS is not protected.  The "outlier"  and "fault" rows report respectively  the accuracy of  NIDS protected by the outlier and fault detection methods. We see that even with protections, there are still some 
decisions that are not accurate. While Byzantine attacks are always detected and the compromised modules removed, the attacker may still have an impact on the decision if it takes too long for the mitigation technique to detect the intruder. From figures \ref{fig3} to \ref{fig8}, we see that the outlier method needs more iterations to detect an intruder. In Table \ref{table2}, the outlier method reports generally a greater number of FPs.

\begin{table}[!htb]
\caption{Accuracy of the NIDS.}
\label{table2}
\centering
 \begin{tabular}{|c |c| c |c | c | c | c |}
\hline
Topology & Size & Mitigation & TP &TN&FP&FN\\
\hline
\multirow{8}{*}{Ring}& \multirow{4}{*}{9} &no attack &466 &520&14&0\\\cline{3-7}
&&no detection &521&0&479&0\\\cline{3-7}
&&outlier &522&404&74&0\\\cline{3-7}
&&fault &456&525&4&15\\\cline{2-7}
& \multirow{4}{*}{25}&no attack &527 &473&0&0\\\cline{3-7}
&&no detection &475&0&525&0\\\cline{3-7}
&&outlier &497&503&58&0\\\cline{3-7}
&&fault &506&489&0&5\\\hline
\multirow{8}{*}{Torus}& \multirow{4}{*}{9} &no attack &493&491&16&0 \\\cline{3-7}
&&no detection &499&0&501&0\\\cline{3-7}
&&outlier &495&438&67&0\\\cline{3-7}
&&fault &478&511&0&11\\\cline{2-7}
& \multirow{4}{*}{25}&no attack &492 &507&1&0\\\cline{3-7}
&&no detection &497&0&503&0\\\cline{3-7}
&&outlier &491&456&53&0\\\cline{3-7}
&&fault &518&450&32&0\\\hline
\multirow{4}{*}{Petersen}& \multirow{4}{*}{10} &no attack &501 &487&12&0\\\cline{3-7}
&&no detection &481&0&519&0\\\cline{3-7}
&&outlier &477&458&65&0\\\cline{3-7}
&&fault &481&516&0&3\\\hline
\multirow{4}{*}{Random}& \multirow{4}{*}{10} &no attack &451 &533&16&0\\\cline{3-7}
&&no detection &485&0&515&0\\\cline{3-7}
&&outlier &503&432&65&0\\\cline{3-7}
&&fault &526&464&0&10\\\hline
 \end{tabular}  
\end{table}

\section{Conclusion}
Local computation of consensus-based distributed applications can be hacked by Byzantine attackers falsifying computed consensus information. Several solutions have been proposed in the literature that address Byzantine attacks on consensus algorithms. We
have adapted two of these solutions, one  from 
model-based fault detection and one from outlier detection to protect a consensus-based network intrusion detection system. Results show that each approach can be used to detect consensus loop disruptions and prevent falsifications of NIDS network traffic assessments. Though preliminary, our results also show significant computational costs of these approaches either in terms of the number of iterations to detect attacks (outlier detection) or in terms of the computational cost of each iteration (model-based detection). This might raise issues for deploying  consensus-based NIDS in suitable environments such as wireless ad hoc networks. 


The work we have presented can be extended to dynamic NIDS network topologies where NIDS modules and network links  enter and leave the network dynamically. 
Dynamically configurable consensus algorithms have been analyzed recently in the control theory literature. Such research should provide ways to cut attackers from the network while satisfying the mathematical assumptions requested during the consensus phase.

Finally, this work could be extended to other attack scenarios such a multiple Byzantine attackers. Model-based fault detection approaches for  multiple Byzantine attackers have been proposed, but their computational cost is high, likely not adaptable in practice to NIDSs.  Nonetheless, the precision and mathematical foundations of model-based fault detection are quite attractive, fundamental research and optimization of current techniques may lead to lowering their computational cost. Outlier detection methods are seeing in the literature as effective approaches to address sophisticated attack scenarios. Experimentation seem the  more appropriate path to evaluate the true value and challenges of outlier detection methods in the context of NIDSs.

\bibliographystyle{IEEEtran}\bibliography{ref}
\end{document}